\documentclass[aip,amsmath,amssymb,preprint]{revtex4-2}
\usepackage{graphicx}
\usepackage{dcolumn}
\usepackage{bm}
\usepackage{graphicx}
\usepackage{bm}
\usepackage{chemformula}
\usepackage[utf8]{inputenc}
\usepackage[T1]{fontenc}
\usepackage{mathptmx}
\usepackage{etoolbox}
\usepackage{xargs}                      
\newcommand{\pmgd}{\texttt{pymatgen-analysis-defects}}

\makeatletter
\def\@email#1#2{%
 \endgroup
 \patchcmd{\titleblock@produce}
  {\frontmatter@RRAPformat}
  {\frontmatter@RRAPformat{\produce@RRAP{*#1\href{mailto:#2}{#2}}}\frontmatter@RRAPformat}
  {}{}
}%
\makeatother

\begin{document}
\title{Simulating Charged Defects at Database Scale}
\author{Jimmy-Xuan Shen}
\affiliation{Materials Science Division, Lawrence Livermore National Laboratory, Livermore, California 94550, U.S.A}
\affiliation{Laboratory for Energy Applications for the Future, Lawrence Livermore National Laboratory, Livermore, California 94550, U.S.A}
\author{Lars Voss}
\affiliation{Materials Engineering Division, Lawrence Livermore National Laboratory, Livermore, California 94550, U.S.A}
\author{Joel B. Varley}
\affiliation{Materials Science Division, Lawrence Livermore National Laboratory, Livermore, California 94550, U.S.A}
\affiliation{Laboratory for Energy Applications for the Future, Lawrence Livermore National Laboratory, Livermore, California 94550, U.S.A}

\date{\today}%
\begin{abstract}
    Point defects have a strong influence on the physical properties of materials, often dominating the electronic and optical behavior in semiconductors and insulators.
    The simulation and analysis of point defects is therefore crucial for understanding the growth and operation of materials especially for optoelectronics applications.
    In this work, we present a general-purpose Python framework for the analysis of point defects in crystalline materials, as well as a generalized workflow for their treatment with high-throughput simulations.
    The distinguishing feature of our approach is an emphasis on a unique, unitcell, structure-only, definition of point defects which decouples the defect definition and the specific supercell representation used to simulate the defect.  
    This allows the results of first-principles calculations to be aggregated into a database without extensive provenance information and is a crucial step in building a persistent database of point defects that can grow over time, a key component towards realizing the idea of a ``defect genome'' that can yield more complex relationships governing the behavior of defects in materials. 
    We demonstrate several examples of the approach for three technologically relevant materials and highlight current pitfalls that must be considered when employing these methodologies, as well as their potential solutions.
\end{abstract}

\maketitle

\section{Introduction}\label{sec:introduction}

Understanding the physical characteristics of point defects (including dopants) in semiconductors and insulators is a crucial analysis tool in materials and semiconductor research since a small number of defects and dopants can significantly change the electrical and optical properties of a given material~\cite{McCluskey2012}.
Since defects are usually formed during the initial growth process, defect properties are difficult to isolate and experiment on. 
Hence, first-principles calculations are often used to analyze these properties.  
Specifically, calculations using density-functional theory (DFT)~\cite{kohn_self-consistent_1965,CGWalle_defects_RMP} have been widely adopted as the standard method.
The analysis of point defects from first principles usually begins with an analysis of the defect thermodynamics by calculating the formation energy of the defect at a given set of chemical conditions.
These formation energies can elucidate which defects are likely to form and different growth conditions and provide information on the doping behavior of these defects~\cite{CGWalle_defects_RMP}.
Recently, more advanced analysis capabilities of quantum recombination rates, for radiative~\cite{Dreyer2020Aug}, and non-radiative recombination~\cite{Wickramaratne2016Oct, Alkauskas2014Aug}, have been developed to help understand the interactions of electrons and photons around point defects.

The computational cost for performing first-principles defect calculations is unusually high due to a combination of factors.
First, the periodic simulation cell containing the defect must be large enough to avoid spurious interactions between the atomic relaxation of the defect and its periodic images.
Second, since the defect represents a localized system embedded in a periodic crystal, the system must be simulated using a level of theory that can accurately describe the electronic structure of both the defect and the host crystal.
This usually means performing hybrid functional calculations in a plane-wave basis set, which is notoriously 10 to 100 times more expensive than traditional local or semi-local functional calculations.

Recently, automated defect analysis tools have been developed to facilitate the simulation and analysis of defects in a high-throughput manner.
Notable among these frameworks are \texttt{pyCDT}~\cite{Broberg2018}, \texttt{pyDefect}~\cite{Kumagai2021}, \texttt{doped}\cite{Mosquera-Lois2023Feb}
and \texttt{DASP}~\cite{Huang2022}.
\texttt{pyCDT}, \texttt{pyDefect} and \texttt{doped} are Python packages aimed at automating the thermodynamic analysis of defects, while \texttt{DASP} also includes the simulation of various quantum recombination rates.

While these tools accomplish their state goal of automating defect analysis, their adoption by the broader defects research community has been limited and building a large-scale persistent defects database is still out of reach. 
This is a key limitation towards the realization of a curated ``Defect Genome'' that can be refined over time and complement the more general Materials Genome Initiative for accelerating materials engineering and discovery.~\cite{Yan2024Jan}
We believe that significant improvements can be made in three areas that will lead to broader adoption: (I) improved integration with existing workflow automation and data aggregation tools, (II) conceptual separation between the code for standard defect analysis and high-throughput calculations, and (III) simulation-independent definition of defects which allows for calculations to be grouped in a database without explicit provenance information.
In this work, we present the advances we have made in these areas and demonstrate how a persistent database of defect properties can be built.

The core software driving our defect analysis is \pmgd{}~\cite{Shen2024Jan}, a part of the \texttt{pymatgen}~\cite{pymatgen} suite materials informatics.
To address (I) and (II), we developed \pmgd{} to be a general-purpose tool for the analysis of defects in crystalline materials and delegated all of the code for high-throughput workflow automation to the \texttt{atomate2} package built upon the \texttt{jobflow} framework for automating general high-performance computing tasks~\cite{Rosen2024Jan}.
This allows us to codify common defect analysis tasks within \pmgd{} and provide a consistent interface for users to perform these tasks without the added complication of high-throughput calculations.
For high-throughput calculations, we have developed a generalized defect workflow as part of the \texttt{atomate2} package which is capable of orchestrating thousands of DFT defect calculations and organizing their results in a standardized format.

Finally, for (III), we formalized a structure-only definition for point defects which allows us to group calculation results without explicit provenance information.
A common critique of high-throughput defect calculations is that structure minimization algorithms can miss the global energy minimum, either due to a lack of local symmetry breaking~\cite{Mosquera-Lois2023Feb} or not finding the correct spin multiplicity~\cite{Xiong2023Mar}.
Having a persistent definition that is independent of the simulation details allows us to revisit previous calculations and find lower energy configurations by using different calculation settings or initial conditions.
This allows us to update any database incrementally as more calculations are performed, such as those that increase the accuracy or sample additional structures.
Over time, users can improve both the breadth of their defects databases --- by including calculations for additional charge state, and the quality of data --- by finding relaxations to local or global energy minima that were previously missed by the automated workflow.

\section{Background and Methods}\label{sec:methods}

The thermodynamic properties of a point defect ($X$) are controlled by the defect formation energy ($E^f[X^q]$), which measures the energy cost of creating a defect with a charge state $q$ at a given set of chemical conditions.
The formation energy is determined by the chemical potentials of everything that goes into forming the defect~\cite{Broberg2018,CGWalle_defects_RMP}:
\begin{equation}\label{eq:formation_energy}
    E^f[X^q] = E_{\rm tot}[X^q] - E_{\rm tot}[{\rm bulk}] + \sum_i n_i \mu_i + qE_{\rm F} + \Delta^q \, ,
\end{equation}
where $E_{\rm tot}[X^q]$ is the total energy of the defect simulation supercell, 
$E_{\rm tot}[{\rm bulk}]$ is the total energy of the bulk cell of the same size; $n_i$ is the change in the number of atoms of type $i$ required to form the defect, while $\mu_i$ is the chemical potential of atom type $i$;  $E_{\rm F}$ is the chemical potential of the electron (often called the Fermi level), and $\Delta^q$ is the finite-size correction required to account for the fact that we are simulating a charged defect in a finite periodic cell.

While all of the different chemical potentials are interconnected, it is conceptually easier to separate them into two categories: 
the chemical potential of the atoms and the chemical potential of electrons. 
The chemical potential of electrons (often called the Fermi level) accounts for how all of the external conditions, including the presence of other defects, affect the energy costs of adding or removing an electron from the defect system. 
Since the electrons in the system can be manipulated after the defect is formed, the Fermi level is often considered a free variable and is shown as the x-axis whenever $E^f[X^q]$ is plotted for a given set of the other chemical potentials.
The chemical potentials of the different atoms added to or removed from the defect are less dynamic and assumed to be fixed after the defect is formed.
So the formation energy of each charge state of a defect is typically shown as a linear function of $E_{\rm F}$ with a slope of $q$ with a $y$-intercept determined by the chemical environment during crystal growth.

The core goal of any defect analysis code is to calculate the formation energy of a given defect in the different charge states.
However, since $E_{\rm tot}[X^q]$ is calculated by creating a point defect in a periodic supercell and $\Delta^q$ presumably eliminates the finite size effects, we can have many different supercell calculations that all correspond to the same defect charge state.
Traditionally, users will have to track the provenance of the defect and charge state for all the calculations.
This can quickly become untenable as the number of defects and charge states increases.
Additionally, since the most stable spin and atomic configuration of a defect might not be found by the first structure relaxation, it is often necessary to revisit defect calculations to improve upon the data over time.
This necessitates a way for us to compare and group defects without explicit provenance information.
Toward that end, we have developed a structure-only definition of point defects that will allow persistent, incrementally growing databases of defect simulation results to be constructed (cf. Sec.~\ref{sec:defect-definition}).
To compute the different atomic chemical potentials ($\mu_i$), we combine explicit DFT calculations of the pure elemental phases and experimentally corrected values of the formation enthalpy to arrive at a more accurate measure of the elemental chemical potentials with fewer calculations.

For the calculation of the finite-size correction term ($\Delta^q$), we use the finite-size correction method of Freysoldt, Neugebauer, and Van de Walle (FNV)~\cite{Freysoldt2009}. 
Previous implementations~\cite{Freysoldt2009,Broberg2018} of this method required the position of the defect to be supplied by the user.
Since we require our data to be provenance-agnostic, we have developed a method to calculate the position of a point defect without prior knowledge of how the defect was defined (cf. Sec.~\ref{sec:position}).
This eliminates a persistent point of user intervention in the workflow of defect analysis and allows for more software-driven high-throughput approaches.

The combinations of these advances allow us to arrive at an analysis framework to define, simulate, and analyze defects in a fully automated manner.
All of the code responsible for the definition and analysis of point defects is available via the \texttt{pymatgen-analysis-defects} add-on package to \texttt{pymatgen}.~\cite{Ong2013Feb}
The automation workflow is distributed independently as part of the \texttt{atomate2} package which allows us to orchestrate and organize thousands of quantum chemistry calculations (cf.~\ref{sec:workflow}).
We will detail these advancements in the remainder of this section.

\subsection{Structure-only Definition of Defect}\label{sec:defect-definition}
For most DFT calculations, the physical system you are simulating is solely defined by the atomic positions and the lattice vectors of the unit cell.
This allows large databases of DFT calculations to be built by simply storing the atomic structures and aggregating all of the calculations related to a given structure into a single entry.
While the unit cell is not unique, modern symmetry analysis tools allow different representations of the same periodic structure to be matched to each other so it is trivial to group calculations that represent the same physical system.
However, when simulating point defects, we are using an arbitrary supercell to simulate the defect in the dilute limit, so multiple symmetry-distinct supercell structures can all correspond to the same point defect.
This inherently breaks the one-to-one mapping between structures and calculations, and thus the structure matching along is not enough facilitate the building of a persistent database of defect calculations.
A core feature of \pmgd{} is the ability to define defects in a structure-only manner, which allows us to group calculations for the same defect without explicit provenance information.
Ultimately, this definition frees us from the need to explicitly track the relationship and dependencies between sets of defect calculations and makes database building significantly easier.

In \pmgd{}, a defect is defined by the combination of:
\begin{itemize}
    \item The type of defect: whether it is a vacancy, interstitial, or substitutional defect.
    \item A unique primitive unit cell of the bulk material.
    \item The fractional position of the defect in the unit cell.
\end{itemize}

\begin{figure}[ht]
    \centering
    \includegraphics[width=0.75\columnwidth]{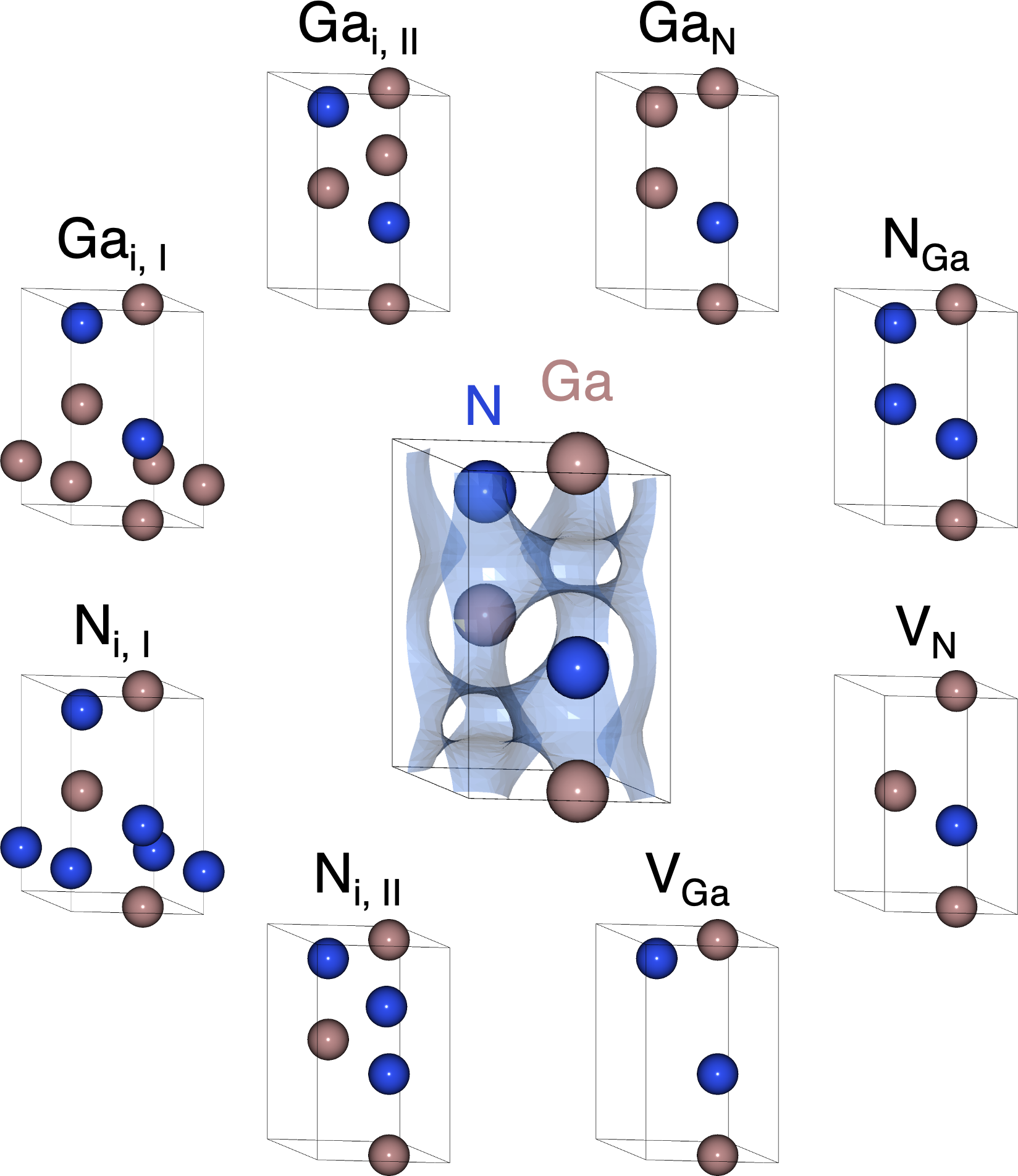}
    \caption{Defect structures for native point defects in GaN, note that two symmetry distinct initial interstitial sites were identified.}
    \label{fig:defect_structs}
\end{figure}

This definition allows us to create a representative structure ($\mathcal{S}_{X}$) of the defect $X$ in the unit cell.
As long as the primitive cell is fixed, two defects $X$ and $Y$ are considered to be the same if and only if $\mathcal{S}_{X}$ and $\mathcal{S}_{Y}$ can be structurally matched to each other, i.e. they are equivalent under the symmetry operations of the primitive cell.
While there might be multiple equivalent sites in a primitive cell for a point defect to form, the structures representing these different versions will be symmetrically equivalent to each other.
As examples, all of the symmetry-distinct native defects of wurtzite GaN are shown in Fig.~\ref{fig:defect_structs}.
The vacancy and antisite defects are trivially constructed, however, generating interstitial defects has always been challenging.
Following the method from Ref.~\onlinecite{Shen2020Oct}, the symmetry-distinct (as defined using \texttt{Spglib}~\cite{Togo2018Aug}) local minima in the charge density were identified as possible interstitial insertion sites, which resulted in two symmetry-distinct versions of the Ga and N interstitial defects. 
Note that this method and be easily combined with charge density API~\cite{Shen2022Oct} from the Materials to quickly generate defect structures without additional DFT calculations.

These representative defect structures can be used to define additional quantities such as predicted oxidation states of the elements and degeneracy factors which are required to calculate various thermodynamic quantities related to point defects~\cite{CGWalle_defects_RMP}.
Additionally, once a defect has been created, we can generate a nearly cubic supercell structure containing that point defect to minimize the effects of periodic images.
The resulting calculations can be associated with the original defect so the thermodynamic quantities from the DFT calculations amount to one representation of a concrete concept such as ``energy of V$_{\rm Ga}$ in the $q=-1$ charge state''.
For the 4-atom primitive cell of GaN, we can create a 128-atom supercell that has orthogonal lattice vectors that differ in length by less than 10\%.
For the remainder of this work, explicit supercell defect DFT calculations will be performed using this GaN supercell structure.

\subsection{Chemical Potential Corrections Based on Experimental Data}\label{sec:chempot}

The chemical potential is a measure of the thermodynamic cost of adding or removing an atom from the defect when the formations of different competing phases are considered.
If the formation of the defect is only competing with the pure elemental phases, then the chemical potential is simply the energy per atom from the DFT calculation of the elemental phase ($\mu^{\rm DFT}_i$).
However, competing compounds with lower formation enthalpy can lower the chemical potential ($\mu_i = \mu^{\rm DFT}_i + \Delta \mu_i$) since they create a more stable environment for the defect.
For a set of atomic species (indexed by $i$), and a set of competing compounds (indexed by $\alpha$) these correction terms must satisfy the following constraints:
\begin{alignat}{2}\label{eq:chempot}
    \Delta \mu_i &\leq 0   &{}\forall{}& \text{ elements } i \, \nonumber\\
    \sum_\alpha n^\alpha_{i}\Delta \mu_i &< \Delta H_{f} (\alpha) \qquad &{}\forall{}& \text{ compounds } \alpha \\
    \sum_\alpha n^{\rm bulk}_{i} \Delta\mu_i &= \Delta H_{f} ({\rm bulk}) \nonumber 
\end{alignat}
Where $n^\alpha_{i}$ is the number of atoms of species $i$ in a formula unit of compound $\alpha$, and the $\Delta H_{f}(\alpha)$'s are the formation enthalpies of one formula unit of the compound $\alpha$.
Eq.~\ref{eq:chempot} defines a manifold in chemical-potential space where you are constrained to the facet representing the host material but bound on different sides by the formation of competing phases.
Since the total energy from DFT can change under different calculation settings, we are usually forced to calculate all of the formation enthalpies using the same level of theory.
However, since the formation energy data from the Materials Project is specifically corrected to fit experimental formation enthalpies~\cite{Wang2021Jul}, we can use the formation enthalpy data from the Materials Project to calculate the chemical potential corrections ($\Delta \mu_i$) only and expect similar if not better results.
Hence, the chemical potential in Eq.~\ref{eq:formation_energy} can be expressed as:
\begin{equation}\label{eq:chempot_atom}
    \mu_i = \mu^{\rm DFT}_i + \Delta \mu^{\rm MP}_i
\end{equation}
where we combine the explicitly calculated energy of the elemental phase in a given level of theory with the formation enthalpy correction due to competing phases from the Materials Project. 
This approach is distinct from other approaches based on correcting inaccuracies in chemical potential references, such as fitted elemental-phase reference energies (FERE)~\cite{Stevanovic2012Mar}, which instead corrects the $\mu^{\rm DFT}_i$ terms based on a consistent calculation framework and thermodynamic databases from which to derive the corrections.
From a database-building perspective, this simultaneously provides a more accurate prediction of relative stability between the competing phases (since we are using experimentally-fitted formation enthalpies) and allows users to build a database using only the DFT energy from the elemental phases alone, which significantly reduces the computational cost.

\subsection{Identify Defect Position With No Prior Knowledge}\label{sec:position}

The total energy of the defective supercell ($E^f[X^q]$ in Eq.~\ref{eq:formation_energy}) refers to the energy of the defect simulation supercell after atomic relaxation.
Since atomic relaxation is often unpredictable, it is difficult to know the exact position of the defect in the supercell afterward.
This is especially problematic when we attempt to calculate the finite-size charge correction for native defects in a supercell since the position of the defect is an input that helps determine the potential alignment contribution~\cite{Freysoldt2009}.
Usually, one can arrive at a decent estimate of the defect position by keeping track of the position during the defect-creating processes.
But for cases where the relaxation of the atomic positions is severe, such as the case of split interstitials and split vacancies, the relaxed defect position will differ significantly from the defect creation site.
Additionally, as we aim to build databases without explicit provenance information, we need to find a way to identify the position of the defect in the supercell after structure relaxation without any prior knowledge of the defect creation site.
In \pmgd{}, we included a method to identify the position of the defect in the supercell after structure relaxation by calculating a site-specific distortion field based on SOAP vectors~\cite{Bartok2013,De2016}.
For each site in the relaxed supercell, we calculate the SOAP vectors ($\bm{v}^{\rm defect}_{\sigma}$) for each site $\sigma$ and compare them with the SOAP vectors of the sites in the pristine bulk structure ($\bm{v}^{\rm bulk}_{\sigma^\prime}$).
The distortion on site $\sigma$ is defined as:
\begin{equation}
    \delta_{\sigma} = 1 - \max_{\sigma^\prime} \left\{  \frac{\bm{v}^{\rm defect}_{\sigma} \cdot \bm{v}^{\rm bulk}_{\sigma^\prime}}{\left|\bm{v}^{\rm defect}_{\sigma}\right| \left|\bm{v}^{\rm bulk}_{\sigma^\prime}\right|} \right\}
\end{equation}
To identify the set of sites with the largest distortions, we first sort the distortion values ($\delta_{\sigma}$) in descending order and cut off the descending series at the largest drop in distortion value.
The position of the defect is then given by taking the average position of the most distorted sites, weighted by their distortion value. 

\begin{figure*}[ht]
    \centering
    \includegraphics[width=\textwidth]{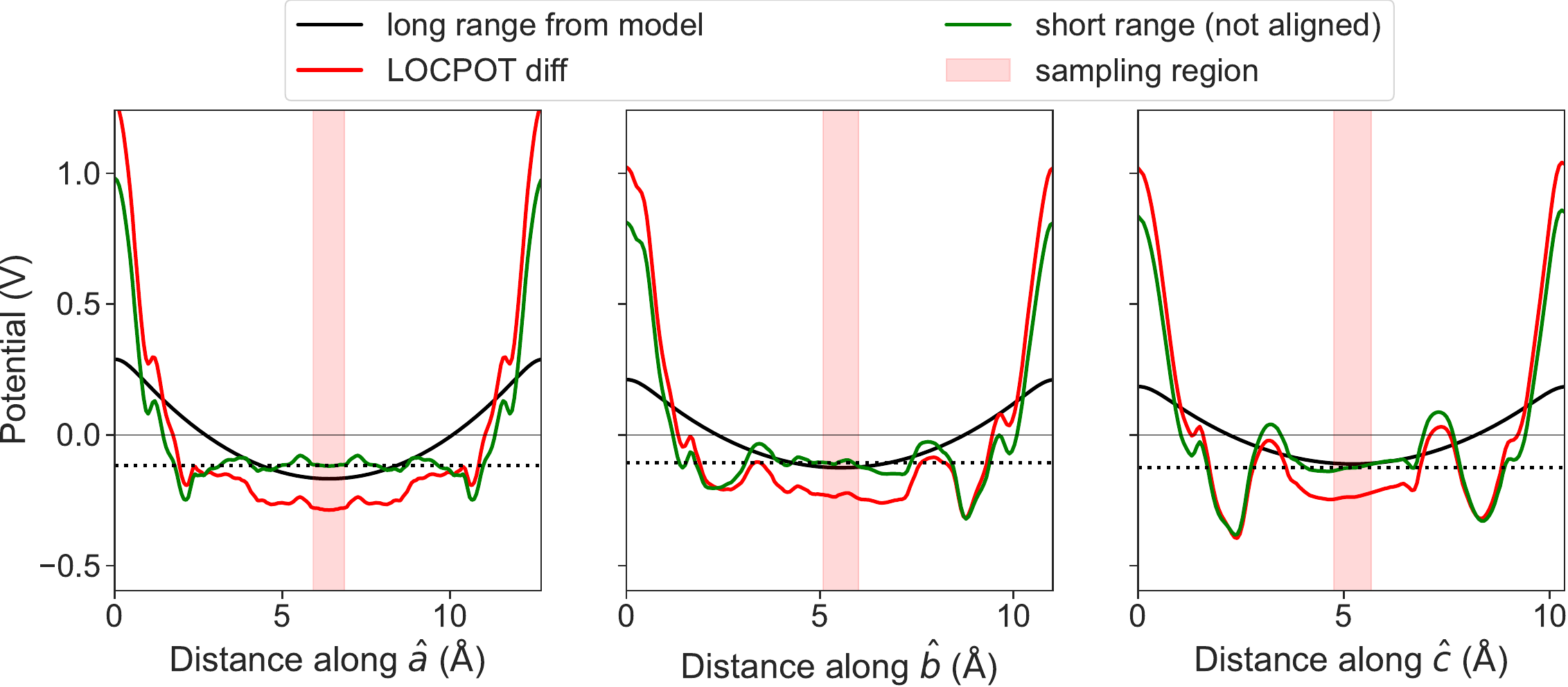}
    \caption{
        Planar-averaged potentials used during the calculations of the FNV charge-state correction for the $V_{\rm Ga}$ defect in GaN.
        The long-range (black) term is computed from a Gaussian model, the potential difference (red) and short-range terms are both shifted so the origin is at the defect positions identified by the finder algorithm.
        The sampling region (shaded red), over which the $C$ offset (dashed) is computed in a small region far from the origin in each crystal direction.
    }
\label{fig:freysoldt}
\end{figure*}

Using this site finder algorithm, we can calculate the FNV finite-size correction with only the electrostatic potentials from the DFT calculations and the dielectric constant of the material.
The alignment-like term in the original Freysoldt correction paper requires the short-range potential ($V^{\rm sr}_{q/0}$) to be shifted by a constant value ($C$) so that the potential is effectively zero far away from the defect.
Fig.~\ref{fig:freysoldt} shows the planar-averaged values of the different potentials involved in calculating the FNV charge-state correction.
The difference in electrostatic potentials between the defect and pristine cells (black curve) and the short-range potential (green curve) are both directly computed from the files storing the electrostatic potential information (e.g. LOCPOT files for VASP calculations).
The position of the defect was automatically determined using the finder algorithm and can be validated by the fact that the potential difference and short-range potential profile both peak at the origin.

This method removed the usual need for user input from previous methods~\cite{Freysoldt2009,Broberg2018} and allow for the entire process of defect formation energy calculation to be codified within our Python framework.
This makes the investigation of singular defects much more approachable and significantly reduces the book-keeping complexity incurred in other high-throughput defect analysis approaches.
Additionally, the ability to easily identify the positions of native vacancies and interstitials in structures using only the atomic positions is likely to have uses beyond the present context.

\subsection{Atomate2 Workflow for Defect Formation Energies}\label{sec:workflow}

The release of \texttt{atomate}~\cite{atomate} and \texttt{aiida}~\cite{Huber2020Sep} marked a major milestone in the open-source materials workflow development, as they codified many standard materials science workflow under a well-defined automation framework.
However, complex workflows such as defect formation energy calculation were difficult to develop due to the fundamental limitations such as processing of the volumetric data and the ability to dynamically create additional DFT calculations based on the results of previous calculations.
\texttt{atomate2} directly addresses these challenges and allows us to define dynamic and composable workflows that can also natively handle the storage of large volumetric data.
With these advances, we were able to develop a flexible and composable high-throughput defect simulation framework that is capable of handling a variety of different simulation needs.

\begin{figure}[ht]
    \centering
    \includegraphics[width=0.85\columnwidth]{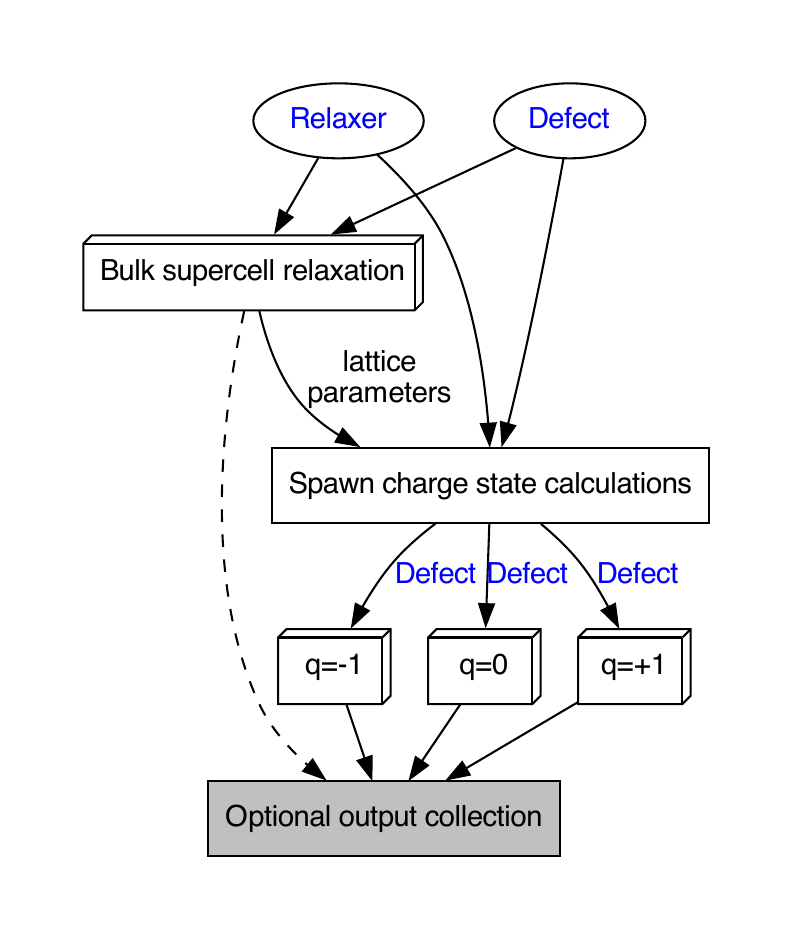}
    \caption{
        Flowchart for the workflow to calculate the defect formation energy. 
        The \texttt{Relaxer} and \texttt{Defect} objects are the required inputs.
        A single bulk supercell calculation is required to obtain the electrostatic potential (e.g. a LOCPOT file for VASP calculations) data and to determine the supercell lattice parameters.
        The outputs from the different charge state supercell calculations can be combined with the bulk calculation to compute the defect formation energy.
        }
\label{fig:workflow}
\end{figure}

The flowchart for the workflow to calculate the defect formation energy of a single defect is shown in Fig.~\ref{fig:workflow}.
The two primary inputs for the workflow are the uniquely defined \texttt{Defect} object (as discussed in Sec.~\ref{sec:defect-definition}) and the \texttt{Relaxer} construct that contains all of the specific calculation settings for the relaxations procedure.
Because we have abstracted away the computational details of the charge state relaxation using the \texttt{Relaxer}, any functional support by VASP~\cite{VASP_ref1} or even entirely different DFT simulation codes (such as CP2K~\cite{CP2K2020}) can be used with minimal modification to the code.

In the workflow for a single defect, we first create a supercell of the bulk material and relax the lattice parameters and atomic positions.
Then, we create a defect supercell with the same lattice parameters as the bulk supercell and relax only the atomic positions of each charge state.
The charge states in this case, are derived from a combination of the allowed formal oxidation states of the species in the bulk formula, and the abundances of specific charge states for these species in the ICSD~\cite{Bergerhoff1983}.
Once these individual charge state relaxations are completed, their raw results are stored in a database along with a serialized version of the original defect object used to generate these calculations.
This allows the final data aggregation step to only examine the \texttt{Defect} objects and the individual calculation settings to quickly find the set of calculations corresponding to charge states of a particular defect simulated using a given level of theory.
The final step is to combine the computed energies and electrostatic potential data from the bulk calculation and different charge state calculations to arrive at a prediction of the formation energy of the defect.
Since we have designed our framework specifically to eliminate the need to track data provenance, this final step can be performed independently from the simulation process.
This becomes more important as a database scales and bookkeeping for the provenance of specific calculations becomes more difficult and mistake-prone.

The composable nature of \texttt{atomate2} allows us to connect many defect workflows to avoid repeating the {\it Bulk supercell relaxation} step.
We can perform this step once and initialize the {\it Spawn charge state calculations} step without any additional DFT calculations.

\section{Results \& Discussions}\label{sec:results}

\begin{figure}
    \includegraphics[width=\columnwidth]{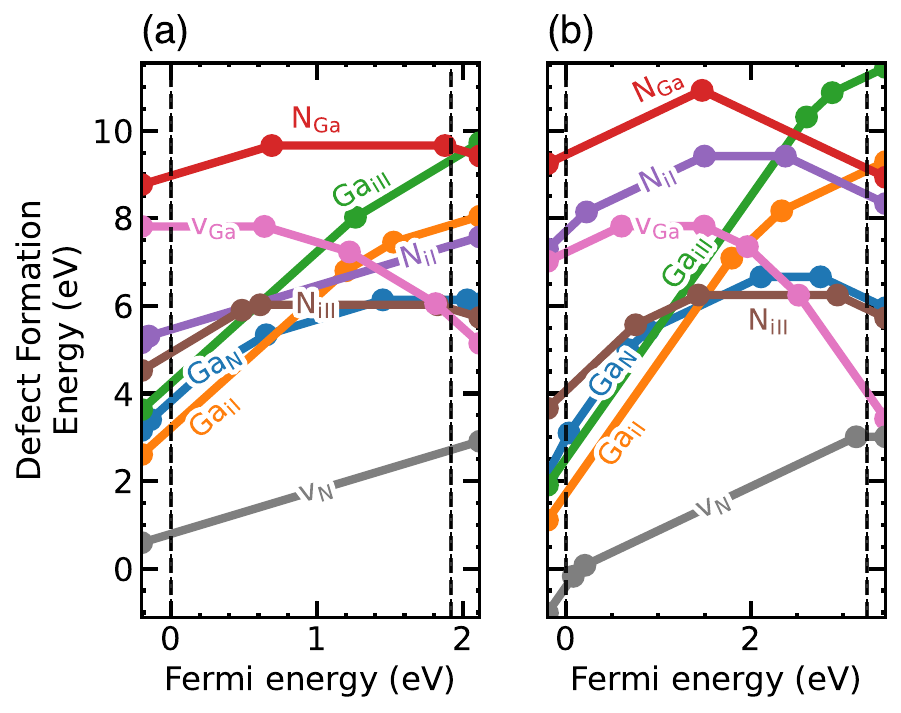}
    \caption{
        Formation energy diagrams for native defects in GaN computed using PBEsol (a) and HSE06 (b). Both sets for formation energy diagrams are calculated using Ga-rich chemical conditions.
        } 
    \label{fig:formation_energy}
\end{figure}

Using the framework described in Sec.~\ref{sec:methods}, we have calculated the formation energies of all of the native defects in GaN at different levels of theory and under different atomic chemical conditions.
The formation energy diagrams for native defects of GaN, computed using HSE06~\cite{Krukau2006Dec} and PBEsol functionals are shown in Fig.~\ref{fig:formation_energy}.
We elected not to fit the exact-exchange fraction to the band gap and used a default of 25\% to make the data directly comparable between material systems.
Nonetheless, the results of our HSE06 calculations match well with previously reported results with a different exact exchange that better represents the experimental band gap of GaN~\cite{Lyons2017Mar} with one notable exception of the N$_{\rm Ga}$ defect.
For N$_{\rm Ga}$, the formation energy of the $q=0$ charge state in HSE06 was too high to be stable relative to the $q=-1$ and $q=+1$ charge states and the estimated charge state range did not agree with the charge states found by Lyons {\it et al.}~\cite{Lyons2017Mar}. 
We use this example to illustrate the first shortcoming in typical high-throughput defect approaches that generate structures that may adopt insufficiently perturbed initial structures or an incomplete set of defect charge states (based on the constituents' oxidation states) that are explored.
The first problem can be fixed by performing the HSE06 calculation with larger distortions to the environment around the defect, which lowered the formation energy of the $q=0$ charge state by 0.58 eV and stabilized the neutral charge state in $n$-type conditions.
The second problem is more subtle and caused by the unusual oxidation behavior of this particular defect.
Since Ga is found in the $+3$ oxidation state and N is found in the $-3$ oxidation state the relevant charge states for the N$_{\rm Ga}$ defect are expected to range from $-6$ to $0$, however, Lyons {\it et al.}~\cite{Lyons2017Mar} found multiple positive charge states for this defect to be stable.
This represents one example of how ``standard" types of point defects may be interpreted as defect complexes for particular oxidation states, e.g. $V_{\rm Ga}$-N$_i$ complexes that can adopt higher positive charge states for Fermi levels closer to the VBM. 
We note that this type of consideration can be factored in {\it a priori} for the prototypical point defect types considered here.
While formal treatment of defect complexes, as well as the integration with informed symmetry-breaking algorithms~\cite{Mosquera-Lois2023Feb} are fully supported by the \pmgd{} framework, they should only be considered on a case-by-case basis due to the increased computational cost.
While these issues only affected a single high-formation-energy defect in those studied in GaN, they highlight the general difficulty in automating defect calculations and why our approach to incrementally updating the data is required. 

We compared our PBEsol and HSE06 calculations and discovered two general methods to improve the overall speed and reliability of defect calculations.
First, we found that the use of a PBEsol calculation to precondition the HSE06 relaxation can significantly reduce the computational cost of the defect calculations. 
However, we acknowledge that this must be considered with care for certain charge states, as a bias toward charge delocalization from semilocal functionals can lead to relaxed structures in local minima which may require additional perturbations or relaxation steps to access global minima for a given defect configuration and charge state~\cite{Mosquera-Lois2023Feb}.
Second, we found that constraining the atoms far from the defect to their bulk positions can both reduce the computational cost and improve the reliability of the defect calculations.
Both of these approaches have been implemented as optional settings that can be easily toggled on or off by the user.

We recommend performing HSE06 defect calculations using a \texttt{Relaxer} that consists of a fast PBEsol relaxation to precondition the structure and wavefunctions, followed by a slower HSE06 relaxation which typically produces more accurate descriptions of the localized defect states and energies. We again stress that the extent to which the defect structures are relaxed with (semi)local functionals before switching to hybrid functionals is sensitive to the particular oxidation state and symmetry of the defect, with electronic closed-shell defects being much more amenable to extended relaxations with semilocal functionals, such as the $V_{\rm Ga}^{-3}$.
For 8 native defects in GaN, we identified a total of 70 distinct charge state calculations that needed to be performed based on our analysis of the formal oxidation state of the defects.
Each electronic optimization step of an HSE06 calculation is usually around 10 to 200 times more expensive than the corresponding PBEsol calculations so we can simply count to number of HSE06 electronic steps taken to determine the relative computational cost of the different approaches.
Using only HSE06 to perform the atomic relaxations, we needed an average of 132.3 electronic steps to complete each charge state calculation.
Using PBEsol atomic to precondition the HSE06 relaxation, we needed an average of 84.2 HSE06 electronic steps to complete each charge state calculation.
This represents a 36\% reduction in the number of electronic steps required with a less than 0.05 eV difference in the final formation energy predictions.

A well-known problem when performing defect calculations is the fact that the perturbation of the crystal caused by the defect can cause another phase transition in the host materials so no matter how large the supercell is, the defect distortion caused by the defect will still cause periodic image interactions.
This can be especially problematic for interstitial defects which can cause large distortions in the surrounding lattice.
These distortions are often not physically meaningful since the bulk phase is stabilized by finite temperature effects not captured by the ground-state DFT calculations.
To avoid this problem, we first calculate the largest sphere that can be inscribed in the supercell and then constrain all atoms outside of this sphere to their bulk positions.
This ensures that the lattice distortions caused by the defect cannot interact with itself across periodic boundaries.

\begin{figure}[ht]
    \includegraphics[width=\columnwidth]{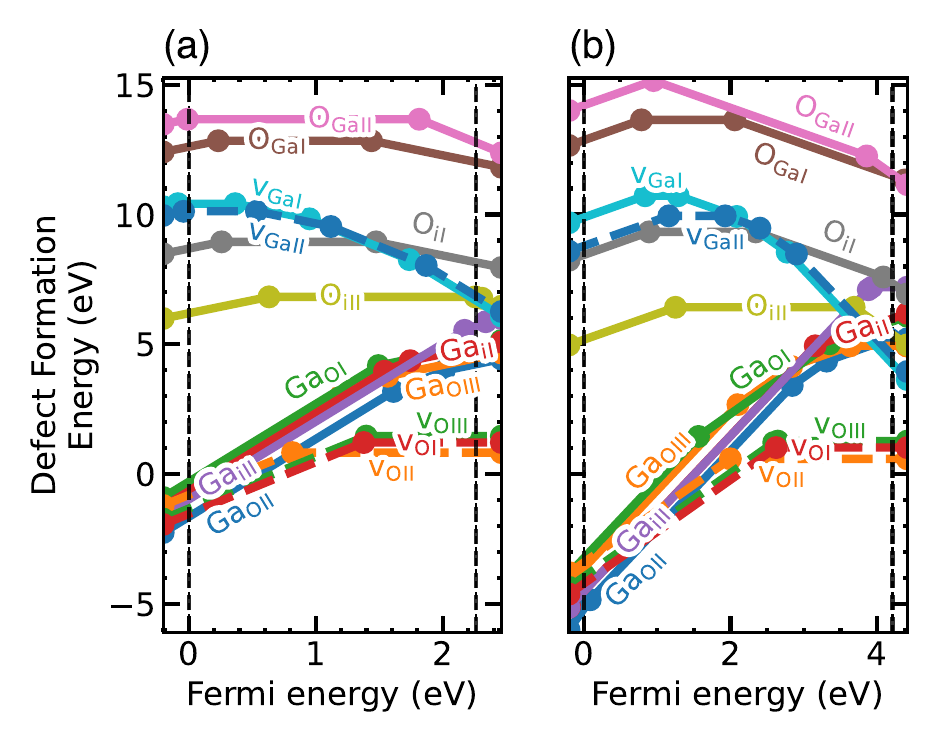}
    \caption{
        Formation energy diagrams for native defects in Ga$_2$O$_3$, computed using PBEsol (a) and HSE06 (b). Both sets for formation energy diagrams are calculated using Ga-rich chemical conditions.
    } 
    \label{fig:formation_energy_Ga2O3}
\end{figure}

Using the same methods for generating and analyzing native defects in GaN, we have also computed the formation energies of the native defects in $\beta$-Ga$_2$O$_3$, a lower-symmetry, more complicated monoclinic structure that has two symmetry-distinct Ga sites and three symmetry-distinct O sites.
This leads to 10 distinct ``simple'' native defects in Ga$_2$O$_3$, excluding interstitials: two each for the $V_{\rm Ga}$ and O$_{\rm Ga}$ defects and three each for the $V_{\rm O}$ and Ga$_{\rm O}$ defects. 
with a number of interstitial sites also possible in the $\beta$-gallia structure.
Using the same charge-density method as above~\cite{Shen2020Oct}, we identified 5 symmetry distinct interstitial sites by analyzing the electronic charge density and they coincide with recently published work on interstitials in Ga$_2$O$_3$.~\cite{Frodason2023Jan}.
Since the number of candidate interstitial sites can be large, we recommend limiting the insertions to the two sites with the lowest average charge density in the surrounding region since that is known to be a good predictor of stability.~\cite{Shen2023Jun}
The formation energy diagrams for native defects in Ga$_2$O$_3$ are again computed with two different levels of theoretical treatment, this time with the PBEsol and HSE06 functionals and shown in Fig.~\ref{fig:formation_energy_Ga2O3}.
As for GaN, we used a standard fraction of exact exchange of 25\% for HSE, which is also known to underestimate the band gap of $\beta$-Ga$_2$O$_3$.
Once again, our fully automated framework largely replicated recently published computational results~\cite{Ingebrigtsen2019Feb} without any human intervention. 
We note that the qualitative orderings of the defects are consistent with other results, and summarize the computed transition levels in the Supporting Information.
However, it has been shown that $\beta$-Ga$_2$O$_3$ exhibits several highly distorted vacancy configurations (so called split-vacancies) that are lower in energy and that the automated relaxation framework was unable to initially capture.~\cite{Varley2011Aug,Ingebrigtsen2019Feb, Kyrtsos2017Jun}
This echoes a key current deficiency in point defect workflows similar to the case with GaN, and highlights the importance of efficient databasing and the ability to apply additional structural perturbations that may be outside the typical tolerance of automated symmetry-breaking algorithms.

As a third case study, we also computed the formation energies of the different native defects of cubic SrTiO$_3$ (STO).
While the cubic perovskite phase is structurally simple, STO thermodynamically prefers a lower-symmetry structure at 0K and also exhibits a more complex chemical stability region given the larger number of elemental constituents. 
The perturbation caused by the defect can lead to spurious transformations to the cubic phase when the defect is simulated in a finite-sized periodic cell.
To alleviate this, we define a sphere centered at the defect with the largest radius allowed by the periodic boundary conditions and freeze all atoms outside this sphere.
This leads to faster convergence of the structural relaxation and more reliable results.

\begin{figure*}[ht]
    \includegraphics[width=\textwidth]{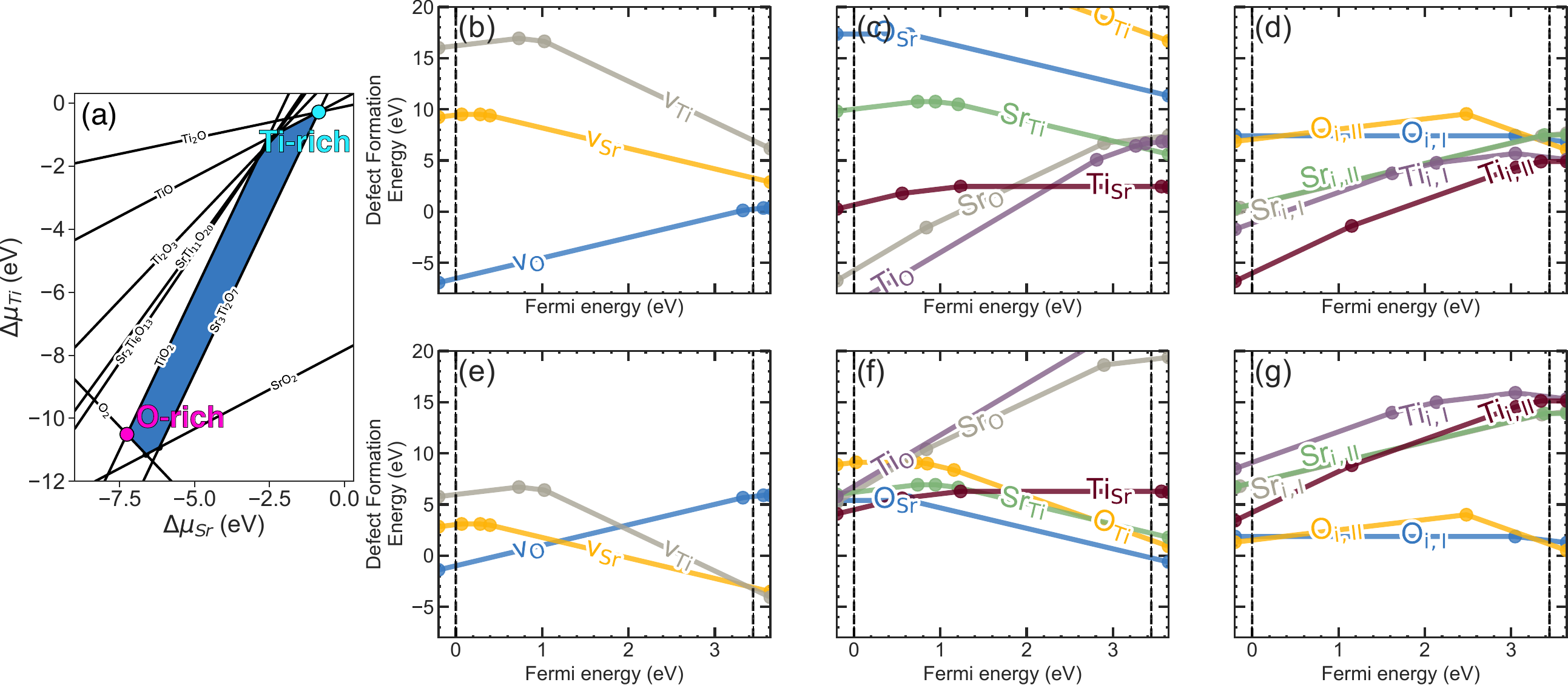}
    \caption{
        (a) The stability region of STO in the (Sr, Ti) chemical potential space, the Ti-rich (teal), and O-rich (magenta) growth conditions are indicated.
        (b-d) The formation energy diagrams of native point defects of STO under Ti-rich conditions.
        (e-g) The formation energy diagrams of native point defects of STO under O-rich conditions.
    } 
    \label{fig:formation_energy_STO}
\end{figure*}

The atomic structure of STO only has a single symmetry-distinct site for each atomic species, resulting in three distinct vacancies and six distinct antisite defects.
Two symmetry-distinct interstitial sites were identified using the electronic charge density, which resulted in another six interstitial native point defects.
Since the Sr-Ti-O chemical space is more complex, the benefit of using thermodynamic databases like the Materials Project to determine the relevant boundaries for the chemical potentials is more apparent.
A total of nine distinct compounds are involved in defining the stability region of STO in chemical potential pace, which would usually require nine additional hybrid DFT calculations on top of the calculations for the total energies of the elemental phases.
Using the method outlined in Sec.~\ref{eq:chempot}, we can skip these nine calculations and use the formation enthalpy data from the Materials Project to compute the chemical potential limits.
The stability region for STO in chemical potential space is shown in Fig.~\ref{fig:formation_energy_STO}~(a), and we have selected the Ti-rich and O-rich growth conditions to represent two limiting conditions for the growth of STO.
This contrasts previous studies~\cite{Liu2014Jul,Baker2017Mar} which only considered \ch{TiO2} and \ch{SrO} as the competing phases and only considered a single interstitial site.

The formation energy diagrams of native defects of STO under O-rich conditions are shown in Fig.~\ref{fig:formation_energy_STO}~(b-d) and the ones for Ti-rich conditions are shown in Fig.~\ref{fig:formation_energy_STO}~(e-g).
Similar to Ref.~\onlinecite{Liu2014Jul} and Ref.~\onlinecite{Baker2017Mar}, we find that the three vacancies $V_{\rm O}$, $V_{\rm Ti}$ and $V_{\rm Sr}$ are the most stable native defects under most conditions.
Under O-rich conditions, the +2 charge state of $V_{\rm O}$ is the most stable defect across most Fermi levels in the band gap.
Under Ti-rich conditions, the $-2$ charge state of $V_{\rm Sr}$ becomes more stable for $n$-type conditions, and the $-4$ charge states of $V_{\rm Ti}$ become stable just below the conduction band.
The relevant defects, charge states, and formation energies of these defects change under different chemical conditions are all consistent with previous studies~\cite{Liu2014Jul,Baker2017Mar} with errors typical for defect calculations.~\cite{Janotti2014Aug,Varley2014Feb,Choi2009Oct}

As we have seen from the examples above using GaN, Ga$_2$O$_3$, and STO, our fully automated framework can reproduce previous computational results without any human intervention.
Instead of a static database of the results at the end, our strategy results in a dynamic database that can be incrementally improved over time as more accurate calculations are performed.
In each example above, we were able to identify a set of defects that are likely to be relevant under a given set of chemical conditions and have a reasonable estimate of the charge state behavior of these defects.

A limitation of any high-throughput approach is the trade-off between generality and accuracy.
Here, we have elected to use a single mixing parameter of 25\% for all calculations, which is a choice that simplifies the comparison of energetics within a given level of theory and computational parameters.
However, a fixed mixing parameter prevents us from using the common strategy for tuning the mixing parameter to match the experimental band gap, a typical choice for more direct comparisons of calculated defect transition levels with experiments. 
Choices of ``incorrect" exact exchange will result in some inaccuracies in the formation energies and defect transition levels that are largely predictable and quantifiable given that localization is qualitatively consistent between the calculations, and unlikely to affect the overall trends in the defect behavior.~\cite{Alkauskas2011Apr} 
This re-emphasizes the need to separate the responsibilities of the defect analysis and high-throughput defect simulation since high-throughput defect analysis is often more concerned with the relative stability of the defects and not the exact thermodynamic properties of a specific defect.
The data shown in Figs.~(4-6) are available as tables in the Supplementary Material.

\section{Conclusion}\label{sec:conclusions}

We have demonstrated a general-purpose codebase (\pmgd{}) for the analysis of point defects in crystalline materials.
Our code enforces a strong connection between a symmetry-distinct defect structure and core concepts in defect analysis, allowing for further extension of the code base in an object-oriented fashion.
Although not discussed in the manuscript, we have also implemented methods for configuration-coordinate diagram analysis as well as tools for simulating radiative and non-radiative recombination at defect centers.
This framework codifies some of the more complex computational workflows in materials informatics and the strong integration with existing analysis and automation software allows for more consistent development and maintenance by the community as the core \texttt{pymatgen} codebase has demonstrated over the past decade.

Since the code is designed for minimal user intervention, we were able to integrate this with existing workflow automation tools to build a fully automated high-throughput defect analysis framework.
We have demonstrated that a fully automated approach can largely reproduce previously published computational results without any human intervention and identified current drawbacks with the current implementation. Specifically, we find that undersampling of global and local minima structures of particular oxidation states, particularly in the case of defect complexes or other defects that exhibit large structural distortions, may be missed in initial searches but can be readily revisited with our emphasis on a robust database-centered representation of point defects. These issues are currently being resolved in future versions that will handle symmetry-breaking and defect complexes in a more rigorous manner.
With more computational resources, it is now possible to build a persistent database of defect properties that can be improved over time as more accurate calculations are performed, although that is beyond the scope of the present work.

\section{Supplementary Material}
The transition levels for GaN, Ga$_2$O$_3$ and SrTiO$_3$ are presented as tables in the Supplementary Material.

\section{Acknowledgements}
This work was performed under the auspices of the US DOE by Lawrence Livermore National Laboratory under contract DE-AC52-07NA27344 and supported by LLNL LDRD funding under project number 22-SI-003. 

\section{Author Declarations}
\subsection{Conflict of Interest}
The authors have no conflicts to disclose.

\subsection{Author Contributions}
JXS contributed to the software development and the manuscript.
JBV contributed to the manuscript.
LV supervised the project and reviewed the manuscript.

\subsection{Data Availability}
The data that supports the findings of this study are available within the article and its supplementary material.
The code used to generate the data is released under the modified BSD license and is available at \url{https://github.com/materialsproject/pymatgen-analysis-defects}.

\bibliography{BIBLIO}

\end{document}